\documentclass[aps, prd, twocolumn, nofootinbib, showpacs]{revtex4}

\usepackage{amsfonts,amsmath,amsthm,amssymb,graphicx,hyperref}

\begin{document}


\title{DBI analog of a decaying vacuum cosmology}

\author{Dennis Bessada${}^{1,2}$,\footnote{
        {\tt dennis.bessada@unifesp.br}}
        }

     \affiliation{
              ${}^1$Laborat\'orio de F\'\i sica Te\'orica e Computa\c c\~ao Cient\'\i fica, Universidade Federal de S\~ao Paulo - UNIFESP, Campus Diadema, Brazil\\
              ${}^2$INPE - Instituto Nacional de Pesquisas Espaciais - Divis\~ao de Astrof\'isica, S\~ao Jos\'e dos Campos, 12227-010 SP, Brazil}

\begin{abstract}

In this work I discuss the dynamical and thermodynamical
equivalence between a general {\it k}-essence scalar field
cosmology and an arbitrary cosmological model with a decaying
vacuum, thus generalizing the approach proposed by Maia and Lima
[Phys. Rev. D {\bf 65}, 083513 (2002)]. The formalism obtained is
quite general and holds for any non-canonical scalar field model.
As a special case I derive a Dirac-Born-Infeld (DBI) model with an
exponential potential and constant speed of sound, and show that
it is equivalent to a cosmological model with decay law
$\Lambda(H) = 3\beta H^2$.

\end{abstract}


\maketitle

\section{Introduction}
\label{sec:introduction}

The discovery of cosmic acceleration \cite{acceleration} opened up
a new research field in cosmology. Many models have been proposed
to account for this observational evidence, and the dark energy
(DE) hypothesis, in which a fluid with negative pressure is
responsible for driving cosmic acceleration, seems to be the most
promising candidate. However, the physical nature of DE is still
obscure. The most accepted paradigm establishes the cosmological
constant $\Lambda$ (CC for short)  as playing the role of such
``fluid" with negative pressure, dominating the energy density of
the universe today. Despite fitting well the available data, this
approach is plagued with some problems (see
\cite{Perivolaropoulos:2008ud} for a discussion); among them, it
is worth recalling the the so-called {\it fine-tuning problem},
(or {\it the old CC problem}) and the {\it coincidence problem}.
The first one is related to the fact that the present-time
observed value for the vacuum energy density,
$\rho_{\Lambda}=\Lambda c^2/(8\pi G)\sim 10^{-47}\,\rm{GeV}^4$, is
more than $100$ orders of magnitude smaller the value found using
the methods of quantum field theory (QFT) ($\sim
10^{71}\rm{GeV}^4$) \cite{Weinberg:1988cp}. The coincidence
problem is related to the fact that the vacuum energy density
started to dominate over the matter energy density just at the
present cosmological time.

A natural attempt to alleviate both problems relies on the
introduction of additional dynamics to the cosmological model in
order to make the vacuum energy density evolve with time; thus,
the corresponding vacuum energy density would have a high enough
value to drive inflation at the very early universe, decaying
along the expansion history to its small value observed today.
Such mechanism can be implemented with the help of scalar fields
\cite{quintessence}, for example; in this case, there is a phase
in which the potential dominates over the kinetic energy, leading
to the desired negative pressure to drive cosmic acceleration.
Also, along with this ``canonical" scenario there is also the
so-called ``noncanonical" approach, in which the scalar field
Lagrangian exhibits a nonlinear kinetic term. This class of
models, also called {\it {\it k}-essence}, appeared first in the
context of inflation \cite{ArmendarizPicon:1999rj}. It was later
generalized to account for the cosmic acceleration
\cite{ArmendarizPicon:2000dh}, and thereafter many proposals have
been done in the literature following this path. Some of them are
derived as low-energy solutions in string theories, and the
possibility of being derived from a more fundamental theory is one
of their attractive features, provided they solve the same puzzles
as the conventional theories do. One of the most promising
string-inspired proposals is the so-called Dirac-Born-Infeld (DBI)
model \cite{Silverstein:2003hf,Alishahiha:2004eh}, which comes out
from a $D3$-brane motion within a warped compactification. DBI
models can provide very interesting inflationary (see, for
example, \cite{Bessada:2009pe} and references therein), as well as
DE-like solutions \cite{Martin:2008xw,Ahn:2009hu,Ahn:2009xd}. In
particular, in DBI inflation the inflaton field is interpreted as
the distance between two branes moving in the extra dimensions
along a warped throat; thus, instead of being inserted
phenomenologically in the Lagrangian, the inflaton actually
emerges from fundamental physics in this scenario.

As an alternative to the scalar field approach, it is also
possible to introduce additional dynamics to the cosmological
model by means of a phenomenological time-dependent cosmological
term $\Lambda(t)$
\cite{vacdecay,Freese:1986dd,Peebles:1987ek,Carvalho:1991ut,Lima:1994ni,Overduin:1998zv}
(see also \cite{Sola:2011qr} for a discussion of $\Lambda(t)$
models arising in the context of QFT in curved space-time). In
this scenario, the time-dependent cosmological term yields a
coupling with another cosmic component, which implies either
particle production or an increase in the time-varying mass of the
dark matter particles \cite{Alcaniz:2005dg}.

Since $\Lambda(t)$ models are essentially of phenomenological
nature, it is of extreme importance to derive them from
fundamental physics to allow, for example, their embedding in
high-energy theories or in modified gravity models as suggested in
DBI models. A first step towards this goal has been achieved in
\cite{Maia:2001zu}, where the authors derived a classical scalar
field model with dynamical and thermodynamical properties
equivalent to those of some particular $\Lambda(t)$ models. Among
other results, they have found an equivalence between a decaying
vacuum model ruled by the law $\Lambda(H) = 3\beta H^2$ and a
scalar field model with an exponential potential. The key to this
connection lies on two fundamental assumptions: first, that both
approaches obey the same dynamical equations parametrized by a
factor $\bar\gamma$, which depends on the specific $\Lambda(t)$
model adopted; second, they follow the same equilibrium
thermodynamical equations (deduced in \cite{Lima:1995kd}) for the
source terms (which represent the particle creation process) and
for the temperature laws. A natural generalization of such methods
would include noncanonical scalar fields, and this is precisely
the main goal of this work. The equations derived here are quite
general, and hold for any noncanonical field; also, I show that
all the results obtained in this paper reduce to those found in
\cite{Maia:2001zu}. As a noncanonical example I choose a DBI model
with a constant speed of sound and exponential potential, and show
that it is dynamical- and thermodynamically equivalent to a
decaying-vacuum model with the law $\Lambda(H) = 3\beta H^2$.

The present paper is organized as follows: in Section
\ref{sec:backeq} I review the basics of cosmological models with
vacuum decay, whereas in Section \ref{sec:thermo} I discuss their
thermodynamical properties. In Section \ref{sec:nonca} I review
the basics of noncanonical scalar field models and set up the
dynamical and thermodynamical equivalence between such class of
models and the decaying vacuum ones. In Section \ref{sec:partcase}
I derive a particular DBI model to illustrate the methods
developed.

\section{Cosmological models with a decaying vacuum}
\label{sec:vacdecay}

\subsection{The background equations}
\label{sec:backeq}

Throughout this paper I consider a flat, homogeneous and isotropic
universe described by the Friedman-Robertson-Walker (FRW) metric
\begin{eqnarray}
\label{eq:frwmetric}
ds^2 = dt^2 - a(t)^2\left(dr^2 + r^2d\theta^2 +
r^2\sin^2\theta\,d\phi^2\right),
\end{eqnarray}
and filled with a perfect fluid with energy density $\rho$ and
pressure $P$, whose stress energy-momentum tensor is given by
\begin{eqnarray}
\label{eq:emtensorm}
T_{m}^{\alpha\beta} = \left(\rho + P\right)u^{\alpha}u^{\beta} - P
g^{\alpha\beta},
\end{eqnarray}
where $u^{\alpha}$ is the fluid four-velocity. In this picture, it
is assumed that the perfect fluid pressure and energy density are
connected via the equation of state
\begin{eqnarray}
\label{eq:eos}
P = w\rho = (\gamma-1)\rho,
\end{eqnarray}
where $\gamma$ is the barotropic index.

In the presence of a cosmological term $\Lambda$, Einstein field
equations read
\begin{eqnarray}
\label{eq:Einsteineq} 
G^{\alpha\beta} - \Lambda g^{\alpha\beta} = \kappa^2
T_{m}^{\alpha\beta},
\end{eqnarray}
where $\kappa^2 \equiv M_P^{-2}\equiv 8\pi G$, $M_P$ being the
reduced Planck mass. Taking the covariant divergence of both sides
of equation (\ref{eq:Einsteineq}), and assuming that the fluid
energy-momentum satisfies the conservation law
${T_{m}^{\alpha\beta}}_{;\beta} = 0$, it follows that the
covariant divergent of $\Lambda$ also vanishes, since, by the
Bianchi identities, ${G^{\alpha\beta}}_{;\beta} = 0$, which sets
$\Lambda$ to be a constant. In this picture, the cosmological
constant $\Lambda$ is a purely geometrical entity; however, if we
move the cosmological term to the right-hand side of the same
equation, and interpreting $\Lambda$ as a second fluid, there is
no further reason to keep this term constant with respect to time.
In this case, we define the {\it effective} energy-momentum tensor
for the two fluids as
\begin{eqnarray}
\label{eq:defteff}
{\bar{T}}^{\alpha\beta} \equiv T_{m}^{\alpha\beta} +
M_P^2{\Lambda}g^{\alpha\beta},
\end{eqnarray}
which naturally satisfies the energy and momentum conservation
constraint
\begin{eqnarray}
\label{eq:emconstraint}
{\bar{T}}^{\alpha\beta}{}_{;\beta} = 0
\end{eqnarray}
as a consequence of the Bianchi identities.

Next, substituting the metric (\ref{eq:frwmetric}) into
(\ref{eq:Einsteineq}), we get the Friedman equations
\begin{eqnarray}
\label{eq:friedman1}
\kappa^2\rho + \Lambda &=& 3H^2,\\
\label{eq:friedman2}
\kappa^2 P - \Lambda &=& - 2\dot{H} - 3H^2,
\end{eqnarray}
where $H=\dot{a}/a$ is the Hubble parameter; also, from the energy
conservation constraint (\ref{eq:emconstraint}) we get the
continuity equation
\begin{eqnarray}
\label{eq:contL} 
\dot\rho + 3 H\left(\rho + P\right)= - M_P^2\dot\Lambda.
\end{eqnarray}

Hence, a cosmological model with varying $\Lambda$ implies that
energy is transferred from this cosmological term to the perfect
fluid; in other words, the vacuum content decays into particles.
Although particle creation usually leads to nonequilibrium
processes, it is also possible to find a particular configuration
of the system in which equilibrium relations still hold, as we
shall discuss in Section \ref{sec:thermo}.

Next, adding up equations (\ref{eq:friedman1}) and
(\ref{eq:friedman2}), we get the expression
\begin{eqnarray}
\label{eq:rpP}
\kappa^2\left(\rho + P\right) = -2\dot{H},
%
\end{eqnarray}
which yields, using equations (\ref{eq:eos}) and
(\ref{eq:friedman1}),
\begin{eqnarray}
\label{eq:eqgeff}
\frac{2}{3}\frac{\dot H}{H^2} = -\bar{\gamma},
\end{eqnarray}
where we have defined \cite{Maia:2001zu}
\begin{eqnarray}
\label{eq:defgeff}
\bar{\gamma}\equiv \gamma\left(1 - \frac{\Lambda}{3H^2}\right).
\end{eqnarray}
The function $\bar{\gamma}$ correlates the time dependence of each
$\Lambda(t)$ model with the Hubble parameter and its derivative.
This fact is very convenient to our purposes, for given a
particular $\Lambda(t)$ model, we can derive the corresponding
expression for $\bar{\gamma}$ by means of equation
(\ref{eq:defgeff}), and use it to solve equation (\ref{eq:eqgeff})
for its scalar-field analog. We shall address this correspondence
in Sections \ref{sec:nonca} and \ref{sec:DBIvacdec},
\ref{sec:partcase}.

In terms of this quantity, Friedman equation (\ref{eq:friedman1})
yields the following expression for the fluid energy density
\begin{eqnarray}
\label{eq:rhogeff}
\rho = 3M_P^2H^2\frac{\bar{\gamma}}{\gamma};
\end{eqnarray}
next, defining the source term for the cosmological ``fluid"
$\Lambda$ appearing in equation (\ref{eq:contL}) as
\begin{eqnarray}
\label{eq:defF} 
F \equiv - M_P^2\dot\Lambda,
\end{eqnarray}
we see that it is related to $\bar{\gamma}$ through
\begin{eqnarray}
\label{eq:defFterm}
F = 3H\gamma\rho\left(1 - \frac{\bar{\gamma}}{\gamma} +
\frac{1}{3H\gamma} \frac{\dot{\bar{\gamma}}}{\bar{\gamma}}\right),
\end{eqnarray}
where we have used (\ref{eq:contL}), (\ref{eq:eqgeff}) and
(\ref{eq:rhogeff}).

\subsection{Thermodynamical properties of $\Lambda$-decaying models}
\label{sec:thermo}

As I have briefly mentioned in the last section, particle creation
usually leads to non-equilibrium processes. However, under some
simple conditions, we can derive a temperature-evolution law for
vacuum-decaying models with solely the equilibrium contribution.
In this section I deal with this issue, following closely
\cite{Lima:1995kd}.

The macroscopic variables that completely describe the
thermodynamical states of a relativistic perfect fluid are given
by the energy-momentum tensor $T_{m}^{\alpha\beta}$, the entropy
current $S^{\alpha}$ and the particle current $N^{\alpha}$.
Introducing the particle number density $n$ and the specific
entropy per particle $\sigma$, the entropy and particle number
density currents are given respectively by
\begin{eqnarray}
\label{eq:defentrocur} 
S^{\alpha}\equiv n \sigma u^{\alpha},
\end{eqnarray}
and
\begin{eqnarray}
\label{eq:defncur} 
N^{\alpha}\equiv n u^{\alpha}.
\end{eqnarray}
The covariant divergence ${N^{\alpha}}_{;\alpha}$ leads to the
{\it balance equation}
\begin{eqnarray}
\label{eq:defbalance} 
\frac{\dot N}{N} = \frac{\dot{n}}{n} + 3H \equiv \Gamma,
\end{eqnarray}
where $\Gamma$ is the particle creation rate in a comoving volume.
As for the entropy current, the second law of thermodynamics
implies that ${S^{\alpha}}_{;\alpha}\geq 0$.

Next, using Gibbs's relation
\begin{eqnarray}
\label{eq:Gibbs} 
nTd\sigma = d\rho - \frac{\rho + P}{n}dn,
\end{eqnarray}
where $T$ is the temperature, and using equations
(\ref{eq:contL}), (\ref{eq:defF}) and (\ref{eq:defbalance}) we
have that
\begin{eqnarray}
\label{eq:sigmadot} 
\dot\sigma = \frac{1}{nT}\left[F - \left(\rho +
P\right)\Gamma\right] ;
\end{eqnarray}
then, from this equation plus ${S^{\alpha}}_{;\alpha} =
\dot{n}\sigma + n\dot{\sigma} + 3Hn\sigma$, it can be shown that
the temperature evolution law for the relativistic fluid reads
\begin{eqnarray}
\label{eq:templawL} 
\frac{\dot T}{T} = \left(\frac{\partial
P}{\partial\rho}\right)_n\frac{\dot n}{n} +
\frac{\dot\sigma}{\left(\partial\rho/\partial T\right)_n}.
\end{eqnarray}

Notice that, if the specific entropy is constant, one has
$\dot\sigma = 0$, so that
\begin{eqnarray}
\label{eq:FGamma} 
F = \gamma\rho \Gamma
\end{eqnarray}
for a fluid obeying the equation of state (\ref{eq:eos}). Also, we
get from (\ref{eq:templawL}) the usual equilibrium law
\begin{eqnarray}
\label{eq:templawLeq} 
\frac{\dot T}{T} = \left(\frac{\partial
P}{\partial\rho}\right)_n\frac{\dot n}{n} ,
\end{eqnarray}
whose integration yields the usual equilibrium relation for the
particle number density
\begin{eqnarray}
\label{eq:nT} 
n(T)\propto T^{1/(\gamma - 1)}.
\end{eqnarray}
Also, in this ``adiabatic" case, substituting equations
(\ref{eq:defbalance}) and (\ref{eq:FGamma}) into the continuity
equation (\ref{eq:contL}), we get
\begin{eqnarray}
\label{eq:rhoT} 
\rho(T)\propto T^{\gamma/(\gamma - 1)}.
\end{eqnarray}

Hence, even in a vacuum-decaying model the usual equilibrium
thermodynamical relations hold if we assume that the specific
entropy is constant. This is valid if we assume 
that the perfect fluid representing the matter or radiation
content of the universe obeys the equation of state (\ref{eq:eos})
and its number particle density and energy density are given by
the equilibrium equations (\ref{eq:nT}) and (\ref{eq:rhoT})
respectively.

We are able now to relate the equilibrium temperature law
(\ref{eq:templawLeq}) to the expression of the source term $F$ in
terms of $\bar\gamma$, equation (\ref{eq:defFterm}). We can do
this by substituting equations (\ref{eq:eos}),
(\ref{eq:defbalance}), and (\ref{eq:FGamma}) in
(\ref{eq:templawLeq}), which yields
\begin{eqnarray}
\label{eq:templawLgamma} 
\frac{\dot T}{T} = -3H\frac{\gamma - 1}{\gamma}\left[\bar\gamma -
\frac{1}{3\gamma H}\frac{\dot{\bar\gamma}}{\bar\gamma}\right].
\end{eqnarray}

\section{Non-canonical scalar field models}
\label{sec:nonca}

In order to generalize the canonical scalar field description
proposed in \cite{Maia:2001zu}, let us introduce some important
concepts concerning noncanonical models, or {\it {\it k}-essence
models} \cite{ArmendarizPicon:1999rj}. Given a scalar field $\phi$
and the {\it canonical kinetic term} $X
\equiv\partial_{\alpha}\phi\,\partial^{\alpha}\phi/2$, a general
{\it k}-essence field theory is characterized by a noncanonical
kinetic term $F(X)$ in its Lagrangian, where $F$ is an arbitrary
function of $X$ (see \cite{Bean:2008ga} and references therein).
The action of a {\it k}-essence field minimally coupled with
gravity and a perfect fluid is then given by
\begin{eqnarray}
\label{eq:fullaction} 
S = \int d^4x\sqrt{-g}\left[\frac{M_P^2}{2} +
{\cal{L}}\left(X,\phi\right) + {\cal{L}}_m\right];
\end{eqnarray}
from this action we can derive the expression of the
energy-momentum tensor for the noncanonical field,
\begin{eqnarray}
\label{eq:defEMtphi}
T_{{\phi}}^{\alpha\beta} = \left(\rho_{\phi} +
P_{\phi}\right)u^{\alpha}u^{\beta} - P_{\phi} g^{\alpha\beta},
\end{eqnarray}
where the field energy density $\rho_{\phi}$ and pressure
$P_{\phi}$ are given by
\begin{eqnarray}
\label{eq:rhophi}
\rho_{\phi}\left(X,\phi\right) &=& 2X{\cal{L}}_X -
{\cal{L}}\left(X,\phi\right),\\
\label{eq:Pphi} 
P_{\phi}\left(X,\phi\right) &=& {\cal{L}}\left(X,\phi\right),
\end{eqnarray}
and
\begin{eqnarray}
u_{\alpha} = \frac{\partial_{\alpha}\phi}{\sqrt{2X}}.
\end{eqnarray}
The subscript ``$X$" denotes a derivative with respect to the
kinetic term $X$. For a homogeneous field $\phi$ and a FRW
background, the energy-momentum tensor for the background and the
{\it k}-essence fluids is given by
\begin{eqnarray}
\label{eq:EMtphif} 
T_{\phi}^{\alpha\beta} &=& \left(\rho_B + \rho_{\phi} + P_B +
P_{\phi}\right)u^{\alpha}u^{\beta} \nonumber \\ &-& \left(P_B +
P_{\phi}\right) g^{\alpha\beta},
\end{eqnarray}
where the subscript ``$B$" stands for the background fluid. I use
such label to distinguish the fluid component quantities in the
noncanonical field description from their $\Lambda(t)$
counterparts. From this tensor we derive the Friedman equations
for this two-fluid model, which read
\begin{eqnarray}
\label{eq:phifriedman1}
\kappa^2\left(\rho_B + \rho_{\phi}\right)  &=& 3H^2,\\
\label{eq:phifriedman2}
\kappa^2 \left(P_B + P_{\phi}\right) &=& - 2\dot{H} - 3H^2;
\end{eqnarray}
the continuity equation comes from the constraint
(\ref{eq:emconstraint}) applied to (\ref{eq:EMtphif}):
\begin{eqnarray}
\label{eq:contphi}
\dot\rho_B + 3H\left(\rho_B + P_B\right) = - \dot\rho_{\phi} -
3H\left(\rho_{\phi}+P_{\phi}\right),
\end{eqnarray}
which can be rewritten as
\begin{eqnarray}
\label{eq:contphimod} 
\dot\rho_B + 3H\left(\rho_B + P_B\right) = {\cal F} ,
\end{eqnarray}
where
\begin{eqnarray}
\label{eq:defFk} 
{\cal F} \equiv - \dot\rho_{\phi} -
3H\left(\rho_{\phi}+P_{\phi}\right).
\end{eqnarray}
Taking the derivative of (\ref{eq:rhophi}) with respect to the
cosmic time $t$, using the identity
\begin{eqnarray}
\label{eq:idtX}
\frac{d}{dt} = \dot X\frac{\partial}{\partial X} +
\sqrt{2X}\frac{\partial }{\partial \phi}
\end{eqnarray}
and the expression for the speed of sound for a {\it k}-essence
field,
\begin{equation}
\label{eq:defspeedofsound} c_s^{2} = \left(1 + 2X\frac{{\cal
L}_{XX}}{{\cal L}_{X}}\right)^{-1},
\end{equation}
we find that the source term for particle creation due to the
decay of the noncanonical field is given by
\begin{eqnarray}
\label{eq:Fphi}
{\cal{F}} &=&  - \frac{\dot X {\cal L}_{X}}{c_s^2} -
\sqrt{2X}\left(2X{\cal L}_{X\phi} + 3H\sqrt{2X}{\cal L}_{X}
\right.\nonumber \\ &-& \left.{\cal L}_{\phi}\right).
\end{eqnarray}

As we have discussed in Section \ref{sec:backeq}, $\bar{\gamma}$
is an important parameter to specify the cosmological dynamics in
terms of a given $\Lambda(t)$ model. In order to find its
noncanonical scalar field analog, we must rewrite the dynamical
equations of such field in terms of the parameter $\bar{\gamma}$;
to do so, we first introduce the equation of state $P_B =
\left(\gamma - 1\right) \rho_B$ for the background fluid, and then
add equations (\ref{eq:phifriedman1}-\ref{eq:phifriedman2}), so
that
\begin{eqnarray}
\label{eq:rhoB}
\rho_B = 3M_P^2H^2\frac{\bar{\gamma}}{\gamma} -
\frac{\left(\rho_{\phi} + P_{\phi}\right)}{\gamma},
\end{eqnarray}
where we have used the definition for $\bar{\gamma}$, equation
(\ref{eq:defgeff}). Next, introducing the total energy density
$\bar{\rho}$ as
\begin{eqnarray}
\label{eq:defrhobar}
\bar{\rho}\equiv  \rho_{\phi} + \rho_B = 3M_P^2H^2,
\end{eqnarray}
and defining the new variable $x$ as
\begin{eqnarray}
\label{eq:defx}
x\equiv \frac{\rho_{\phi} + P_{\phi}}{\bar{\rho}\bar{\gamma}},
\end{eqnarray}
we see from (\ref{eq:rhoB}) that
\begin{eqnarray}
\label{eq:rhoB1}
\rho_B = \frac{\bar{\rho}\bar{\gamma}}{\gamma} (1 - x).
\end{eqnarray}

The quantity $x$ defined in (\ref{eq:defx}) generalizes its
canonical analog introduced in \cite{Maia:2001zu}. In order to
understand its physical meaning, since it is going to play an
important role in our discussion, let us rewrite this term in the
following way: we add again equations (\ref{eq:phifriedman1}) and
(\ref{eq:phifriedman2}) and use (\ref{eq:defgeff}) and
(\ref{eq:defrhobar}), so that
\begin{eqnarray}
\label{eq:friedman} 
\bar{\rho} + \bar{P} = -2M_P^2 \dot{H} = \bar{\rho}\bar{\gamma},
\end{eqnarray}
where we have defined $\bar{P}\equiv P_B + P_{\phi}$. Hence, in
this formulation the $x$ parameter assumes the form
\begin{eqnarray}
\label{eq:xgen} 
x = \frac{\rho_{\phi} + P_{\phi}}{\bar{\rho} + \bar{P}};
\end{eqnarray}
thus, it measures the relative weight of the noncanonical scalar
field energy density and pressure contribution with regard to the
total energy density of the universe. Note that if $x=0$, one has,
from equations (\ref{eq:rhophi},\ref{eq:Pphi}),
\begin{eqnarray}
\label{eq:xeqzero} 
\rho_{\phi} + P_{\phi} = 2X{\cal L}_X = 0,
\end{eqnarray}
which leads to $X=0$. Hence, in this case, $P_B = -\rho_B$,
recovering the $\Lambda(t)$ scenario. If $x = 1$, we have $\rho_B
= 0$, which leads to an universe whose evolution is driven solely
by a {\it k}-essence field.

Notice that this result is similar to the one obtained in
\cite{Maia:2001zu}; actually, by setting ${\cal
L}\left(X,\phi\right) = X - V$ in (\ref{eq:fullaction}) (which
corresponds to the canonical limit), we get equation (22) in that
reference, as expected.

We next turn to the thermodynamical analysis of the particle
creation for the {\it k}-essence case. As pointed out by
\cite{Maia:2001zu}, the parameter $x$ plays an important role in
the thermodynamics of vacuum decay; for instance, the source term
(\ref{eq:defFk}) reads
\begin{eqnarray}
\label{eq:Fx} 
{\cal{F}} = 3H\gamma\rho_B\left[1 - \frac{\bar{\gamma}}{\gamma} +
\frac{1}{3H\gamma}\left( \frac{\dot{\bar{\gamma}}}{\bar{\gamma}} -
\frac{\dot{x}}{1-x}\right)\right],
\end{eqnarray}
where we have used equations (\ref{eq:contphimod}) and
(\ref{eq:rhoB1}) and the equation of state for the background
fluid. Note that this expression for the source term as a function
of the $x$ parameter appears to be the same as obtained in
equation (29) of \cite{Maia:2001zu}; however, such resemblance is
solely functional, for the differences show up when we write down
the specific expression for $x$ in terms of $\rho_{\phi}$ and
$P_{\phi}$, which are model dependent.

Since we are assuming that the {\it k}-essence field decays into
particles as seen from the continuity equation (\ref{eq:contphi}),
it is important to place the usual thermodynamical constraints on
such process. As in Section \ref{sec:thermo}, we introduce the
corresponding entropy and number density currents for the
noncanonical field decay into matter particles as
\begin{eqnarray}
\label{eq:defentrocurphi} 
{\cal S}^{\alpha}\equiv \nu \,\varsigma u^{\alpha},
\end{eqnarray}
and
\begin{eqnarray}
\label{eq:defnucur} 
{\cal N}^{\alpha}\equiv \nu u^{\alpha},
\end{eqnarray}
where $\nu$ is the particle number density and $\varsigma$ is the
specific entropy for this process. As in (\ref{eq:defbalance}),
the balance equation for the noncanonical case comes from the
covariant divergence ${{\cal N}^{\alpha}}_{;\alpha}$,
\begin{eqnarray}
\label{eq:defbalancephi} 
\frac{\dot{\cal N}}{{\cal N}} = \frac{\dot{\nu}}{\nu} + 3H \equiv
\bar{\Gamma},
\end{eqnarray}
where now $\bar{\Gamma}$ is the particle creation rate in a
comoving volume due to the noncanonical scalar field decay. Next,
substituting equations (\ref{eq:Gibbs}), (\ref{eq:contphimod})
into and (\ref{eq:defbalancephi}) into ${{\cal
S}^{\alpha}}_{;\alpha}$ one gets a similar equation for the
specific entropy time evolution,
\begin{eqnarray}
\label{eq:sigmadotphi} 
\dot\varsigma = \frac{1}{\nu{\cal T}}\left[{\cal F} -
\gamma\rho_B\bar\Gamma\right] ,
\end{eqnarray}
where ${\cal T}$ is the temperature associated with the decay.
Assuming that the laws (\ref{eq:nT}) and (\ref{eq:rhoT}) also hold
for the background fluid, the specific entropy $\varsigma$ is also
conserved in the noncanonical description. Hence, from
(\ref{eq:sigmadotphi}),
\begin{eqnarray}
\label{eq:FGammaphi} 
{\cal F} = \gamma\rho_B \bar\Gamma.
\end{eqnarray}

In this ``adiabatic" case, the temperature law is derived in the
same way as done in Section \ref{sec:thermo}, and is given by
\begin{eqnarray}
\label{eq:templawphi} 
\frac{\dot {\cal  T}}{{\cal  T}} = \left(\gamma -
1\right)\frac{\dot \nu}{\nu},
\end{eqnarray}
which, in terms of the $\bar\gamma$ parameter, reads
\begin{eqnarray}
\label{eq:templawphigamma} 
\frac{\dot {\cal T}}{{\cal T}} = -3H\frac{\gamma -
1}{\gamma}\left[\bar\gamma - \frac{1}{3\gamma
H}\left(\frac{\dot{\bar\gamma}}{\bar\gamma} - \frac{\dot x}{1 -
x}\right)\right].
\end{eqnarray}

Expressions (\ref{eq:Fx}) and (\ref{eq:templawphigamma}) contain
an extra $\dot x$ contribution absent in the $\Lambda (t)$
scenario as seen from (\ref{eq:defFterm}) and
(\ref{eq:templawLgamma}). If we set $x = const.$, we see that the
terms inside the brackets in equations (\ref{eq:defFterm}) and
(\ref{eq:FGammaphi}) are the same, so that
\begin{eqnarray}
\label{eq:Grates} 
\frac{ {\cal  F}}{\rho_B} = \frac{ F}{\rho},
\end{eqnarray}
which implies
\begin{eqnarray}
\label{eq:GratesGamma} 
\Gamma = \bar\Gamma.
\end{eqnarray}
The equality above demonstrates that the ``adiabatic" condition
provides the equivalence mechanism of particle production: the
decay of the vacuum is the same process as the decay of the
noncanonical field.

Next, comparing the temperature laws (\ref{eq:templawLgamma}) and
(\ref{eq:templawphigamma}) for $x = const.$, we see that the terms
inside brackets are also equal, thus
\begin{eqnarray}
\label{eq:Trates} 
\frac{ \dot {\cal  T}}{{\cal  T}} = \frac{ \dot
T}{T}\Longrightarrow {\cal T} = b T,
\end{eqnarray}
where $b$ is a proportionality constant that can be determined as
follows: from (\ref{eq:rhogeff}) and (\ref{eq:rhoB1}) we see that
\begin{eqnarray}
\label{eq:rBr} 
\rho_B = \rho\left(1 - x\right),
\end{eqnarray}
and from (\ref{eq:rhoT}) we obtain
\begin{eqnarray}
\label{eq:TandT} 
\frac{\rho_B}{\rho} = \left(\frac{{\cal
T}}{T}\right)^{\gamma/(\gamma - 1)} = \left(1 - x\right),
\end{eqnarray}
so that $b = \left(1 - x\right)^{(\gamma - 1)/\gamma}$, which
yields the following connection between the temperatures of both
scenarios:
\begin{eqnarray}
\label{eq:TT} 
{\cal T} = T \left(1 - x\right)^{(\gamma - 1)/\gamma}.
\end{eqnarray}

The particle number densities $n$ and $\nu$ can be related by
using (\ref{eq:nT}) and (\ref{eq:TT}):
\begin{eqnarray}
\label{eq:nun} 
\nu = n \left(1 - x\right)^{1/\gamma }.
\end{eqnarray}

As for the entropies of both scenarios, the ``adiabatic" condition
implies, for $\sigma = S/N$ and $\varsigma = {\cal S} / {\cal N}$,
that
\begin{eqnarray}
\label{eq:entropies} 
\frac{\dot{N}}{{N}} = \frac{\dot{\cal S}}{{\cal S}} =
\frac{\dot{S}}{{S}}.
\end{eqnarray}

The assumption $x = const.$ brings about two important
consequences to the approach developed in this paper: first, it
demonstrates that the $\Lambda(t)$ and the {\it k}-essence
scenarios are equivalent from the thermodynamical point of view.
Also, they have equivalent dynamics, for the details of the vacuum
decay, expressed by $\bar\gamma$, also takes part of the dynamics
of the corresponding noncanonical field.

Second, the canonical and all other noncanonical models have the
same source term and obey the same temperature law, provided they
decay into the same background fluid. Hence, all the different
noncanonical models are also thermodynamically equivalent. This
holds because, regardless of the functional form of $\rho_{\phi}$
and $P_{\phi}$, which are model dependent, the sum $\rho_{\phi} +
P_{\phi}$ will be always proportional to $\gamma\rho_B$:
\begin{eqnarray}
\label{eq:xconstant} 
\rho_{\phi} + P_{\phi} = \frac{x}{1-x}\gamma\rho_B.
\end{eqnarray}

Notice that all the results obtained here for an {\it arbitrary}
noncanonical scalar field generalize those found in reference
\cite{Maia:2001zu}, since a canonical scalar field is the simplest
nontrivial particular case of noncanonical models.

\section{Implementing a DBI model with vacuum decay}
\label{sec:dbi}

\subsection{The general theory}
\label{sec:DBIvacdec}

As an application of the above formalism I specialize to a DBI
model. Originally, the DBI Lagrangian emerged from the context of
warped D-brane inflation, in which the inflationary mechanism is
regarded as the motion of a D3-brane in a six-dimensional
``throat" characterized by the metric \cite{klebanov2000}
\begin{equation}
\label{eq:DBImetric} ds^2_{10} = h^2\left(r\right) ds^2_4 +
h^{-2}\left(r\right) \left(d r^2 + r^2 ds^2_{X_5}\right),
\end{equation}
where $h$ is the warp factor, $X_5$ is a Sasaki-Einstein
five-manifold which forms the base of the cone, and $r$ is the
radial coordinate along the throat. In this case, the inflaton
field $\phi$ is identified with $r$ as $\phi = \sqrt{T_3} r$,
where $T_3$ is the brane tension. The dynamics of the D3-brane in
the warped background (\ref{eq:DBImetric}) is then dictated by the
DBI Lagrangian
\begin{equation}
\label{eq:DBIlagrangian} {\cal L} = - f^{-1}\left(\phi\right)
\sqrt{1 - 2 f\left(\phi\right) X } - f^{-1}\left(\phi\right) -
V\left(\phi\right),
\end{equation}
where $f^{-1}(\phi)=T_3h(\phi)^4$ is the inverse brane tension and
$V(\phi)$ is an arbitrary potential. From the DBI Lagrangian
(\ref{eq:DBIlagrangian}) the speed of sound is given by
\begin{equation}
\label{eq:DBIspeedofsound}
c_s(\phi) = \sqrt{1 - 2f(\phi)X},
\end{equation}
where I have used (\ref{eq:defspeedofsound}), whereas the energy
density and pressure for the DBI field are given by
\begin{eqnarray}
\label{eq:DBIrho} 
\rho_{\phi} &=& \frac{1-c_s}{c_sf} + V,\\
\label{eq:DBIP} 
P_{\phi} &=& \frac{1-c_s}{f} - V,
\end{eqnarray}
respectively, derived from equations (\ref{eq:rhophi}) and
(\ref{eq:Pphi}). Next, adding up expressions (\ref{eq:DBIrho}) and
(\ref{eq:DBIP}), and using (\ref{eq:DBIspeedofsound}), we find
that
\begin{eqnarray}
\label{eq:DBIsum} 
\rho_{\phi} + P_{\phi} = \frac{2X}{c_s}.
\end{eqnarray}
The potential can also be read from equations (\ref{eq:rhophi})
and (\ref{eq:Pphi}):
\begin{eqnarray}
\label{eq:DBIV} 
V(\phi) = \frac{c_s\rho_{\phi} - P_{\phi}}{1 + c_s}.
\end{eqnarray}

Such equations hold for any DBI model; thus, in order to write
them as an analog of a specific $\Lambda(t)$ scenario, we must
rewrite such equations in terms of the $x$ parameter, defined by
expression (\ref{eq:defx}). First, using (\ref{eq:defrhobar}) and
(\ref{eq:rhoB1}), we find that
\begin{eqnarray}
\label{eq:DBIrhox} 
\rho_{\phi} = \bar{\rho} + \frac{\bar{\rho}\bar{\gamma}}{\gamma}(x
- 1),
\end{eqnarray}
and from equations (\ref{eq:defx}), (\ref{eq:DBIrho}) and
(\ref{eq:DBIsum}) we get the expression for the pressure of the
noncanonical field,
\begin{eqnarray}
\label{eq:DBIPx} 
P_{\phi} = \bar{\rho}\left[ -1 + \bar{\gamma}x +
\frac{\bar{\gamma}}{\gamma}(1 - x) \right].
\end{eqnarray}

Substituting (\ref{eq:DBIsum}) into (\ref{eq:defx}), we obtain the
following equation for the kinetic term $X$:
\begin{eqnarray}
\label{eq:XDBI} 
X = \frac{3}{2}M_P^2H^2\bar{\gamma}c_s\,x,
\end{eqnarray}
whereas the potential comes from equations (\ref{eq:DBIV}),
(\ref{eq:DBIrhox}) and (\ref{eq:DBIPx}):
\begin{eqnarray}
\label{eq:DBIpot} 
V(\phi) = \bar{\rho}\left[ 1 - \bar{\gamma}\left( \frac{x}{1+c_s}
+ \frac{1 - x}{\gamma}\right) \right].
\end{eqnarray}

Using the definition for $X$, equation (\ref{eq:XDBI}) can be
integrated to find the $a$ dependence of the noncanonical field
$\phi$:
\begin{eqnarray}
\label{eq:eqphi} 
\phi - \phi_{\ast} = \sqrt{3}M_P\int^{a}_{a_{\ast}}
\sqrt{\bar{\gamma}c_s\,x}\,\frac{da}{a},
\end{eqnarray}
where $\phi > \phi_{\ast}$. The integral on the right-hand side of
equation (\ref{eq:eqphi}) depends on the details of the underlying
DBI model and on the explicit time dependence of $\Lambda$ due to
the presence of $\bar{\gamma}$. Then, once fixed a decaying-vacuum
model, there are many DBI descriptions that lead to an equivalent
dynamics for the universe. Among the great variety of possible
choices, I pick here the simplest one, based on the assumption
that both $\bar{\gamma}$ and $c_s$ are constant. Since $x$ is
constant, because I want to assure that the underlying
$\Lambda(t)$ and its corresponding DBI analog are
thermodynamically equivalent, the three constant parameters $x$,
$\bar{\gamma}$ and $c_s$ allow for a straightforward
reconstruction of the functions $f(\phi)$ and the potential
$V(\phi)$ appearing in the DBI Lagrangian
(\ref{eq:DBIlagrangian}). We tackle this particular example in the
next section.

\subsection{A particular case: $c_s$ constant and $\Lambda = 3\beta H^2$}
\label{sec:partcase}

Among the many proposals found in the literature to implement a
vacuum-decaying model (see, in particular, the Table I in
\cite{Overduin:1998zv} for some examples), in this paper I
specialize to the model characterized by the following
phenomenological expression for $\Lambda(t)$ (see
\cite{Freese:1986dd}, \cite{Carvalho:1991ut},
\cite{Overduin:1998zv} and references therein):
\begin{eqnarray}
\label{eq:defLt} 
\Lambda(H) \equiv 3\beta H^2,
\end{eqnarray}
where $\beta\in[0,1]$ is a dimensionless constant parameter. From
(\ref{eq:defgeff}) and (\ref{eq:defLt}) it is straightforward to
show that such law yields $\bar{\gamma} = \gamma\left(1 -
\beta\right)$, which is a constant, as desired; next, I set $c_s$
in (\ref{eq:DBIspeedofsound}) to be a
constant\footnote{Non-canonical scalar field models with constant
speed of sound have already been studied in the literature. It is
worth recalling that a DBI inflationary model with constant (and
low) speed of sound has been discussed in \cite{Spalinski:2007un};
also, another {\it k}-essence model with constant speed of sound
different from unity has also been discussed in
\cite{ArmendarizPicon:1999rj}}. Then, with such constant
parameters, the integral (\ref{eq:eqphi}) can be easily carried
out, and the result is
\begin{eqnarray}
\label{eq:aphi} 
a\left(\phi\right) = a_{\ast} e^{b\left(\phi -
\phi_{\ast}\right)},
\end{eqnarray}
where
\begin{eqnarray}
\label{eq:defb} 
b\equiv \frac{1}{\sqrt{3}M_P\sqrt{\bar{\gamma}\,x\,c_s}}.
\end{eqnarray}

From the definition of $\bar\gamma$, equation (\ref{eq:defgeff}),
we can rewrite the integral (\ref{eq:eqphi}) in terms of $H$,
which yields
\begin{eqnarray}
\label{eq:Hphi} 
H\left(\phi\right) = H_{\ast}
\exp\left[-\frac{3}{2}b\bar\gamma\left(\phi -
\phi_{\ast}\right)\right].
\end{eqnarray}

Combining equations (\ref{eq:aphi}) and (\ref{eq:Hphi}), we see
that the Hubble parameter evolves as a power law of $a$,
\begin{eqnarray}
\label{eq:Ha} 
H\left(a\right) =
H_{\ast}\left(\frac{a}{a_{\ast}}\right)^{-\frac{3}{2}\bar\gamma}.
\end{eqnarray}

From expressions (\ref{eq:DBIspeedofsound}), (\ref{eq:XDBI}) and
(\ref{eq:Hphi}) we find the corresponding expression for the
function $f\left(\phi\right)$:
\begin{eqnarray}
\label{eq:fphi} 
f\left(\phi\right) = \frac{1 - c_s^2}{3M_P\bar\gamma x\, c_s
H_{\ast}}\,e^{3b\bar\gamma\left(\phi - \phi_{\ast}\right)}.
\end{eqnarray}

Lastly, from equations (\ref{eq:defrhobar}), (\ref{eq:DBIpot}) and
(\ref{eq:Hphi}) the potential can be easily evaluated, and yields
\begin{eqnarray}
\label{eq:Va} 
V\left(\phi\right) = V_{\ast}e^{-3b\bar\gamma\left(\phi -
\phi_{\ast}\right)},
\end{eqnarray}
where
\begin{eqnarray}
\label{eq:Vast} 
V_{\ast}\equiv 3M_P^2H_{\ast}^2\left[ 1 - \bar{\gamma}\left(
\frac{x}{1+c_s} + \frac{1 - x}{\gamma}\right) \right].
\end{eqnarray}

\newpage

Therefore, a DBI model with exponential potential and constant
speed of sound is dynamical- and thermodynamically equivalent to a
vacuum-decaying model obeying the law $\Lambda(H) = 3\beta H^2$.
Notice that the canonical result obtained in reference
\cite{Maia:2001zu} is very similar to the one I derived here; the
basic difference concerning the dynamical equations in both models
is the presence of the speed of sound $c_s$, which is equal to one
in the canonical version. Also, unlike the canonical results, in
which $\bar{\gamma}$ and $x$ are the two free parameters to be
constrained by observations, in this DBI model there is another
free parameter, $c_s$, which can take either a small value (less
than one) or even be superluminal\footnote{A solution with $c_s=1$
yields $x=0$ as seen from equation (\ref{eq:xgen}), which
corresponds to the original $\Lambda(t)$ scenario.}. Only a
statistical analysis can constrain this space of free parameters,
and this is the subject of a future work.

\section{Conclusions}
\label{sec:conc}

In this paper I generalize the approach devised in
\cite{Maia:2001zu} to the case of noncanonical scalar fields.
Following the first-order thermodynamical formalism developed to
describe the decay mechanism of the canonical scalar field into
particles \cite{Lima:1995kd}, I derive the corresponding
nonequilibrium relations for the noncanonical case, and obtain
precisely the same equations, showing that noncanonical models are
all thermodynamically equivalent among themselves and to an
arbitrary $\Lambda(t)$ model by assuming that the specific entropy
associated with each cosmic component is conserved (the
``adiabatic" condition). Also, I find the general expression for
the $x$ parameter, responsible for such thermodynamical
equivalence.

As an illustration of the general procedure developed in this work
to deal with noncanonical fields, I specialize to a DBI model with
constant speed of sound. The formalism allows for a potential
reconstruction, and I obtain a DBI model with exponential
potential which is dynamical and thermodynamically equivalent to a
decaying-vacuum model with the law $\Lambda(H) = 3\beta H^2$. In
this particular case there are three parameters to be constrained
by observations: the speed of sound associated with the
noncanonical field, $c_s$, the $x$ and $\bar{\gamma}$ parameters.
The statistical test to constrain such three parameters is the
subject of a future work.

\newpage

\begin{acknowledgments}

I thank Oswaldo Duarte Miranda and Jos\'e Ademir Sales Lima for a
critical reading of this paper.

\end{acknowledgments}

\end{document}